\documentclass[11pt,a4paper,onecolumn,reqno]{amsart}
\usepackage[a4paper, margin=2.3cm, bmargin=3cm]{geometry}

\usepackage{graphicx,stfloats} \usepackage{color}
\usepackage{amsmath}
\usepackage{amsaddr}
\usepackage{bm, soul}
\usepackage{mathtools}
\usepackage[colorinlistoftodos,prependcaption,textsize=tiny]{todonotes}
\usepackage{braket}
\usepackage{hyperref}
\usepackage[square,comma,sort&compress, numbers]{natbib}
\usepackage{enumerate}
\usepackage[version=4]{mhchem}

\usepackage{placeins}

\renewcommand{\st}[1]{}

\usepackage{siunitx}
 
\usepackage{multirow}

\usepackage{etoolbox}
\patchcmd{\pprintMaketitle}
  {\hrule\vskip12pt}
 {\hrule\vskip12pt\ifvoid\extrainfobox\else\unvbox\extrainfobox\par\vskip12pt\fi}
 {}{}

\newsavebox\extrainfobox

\usepackage{caption}
\renewcommand{\figurename}{Fig.}
\title{Effect of flame retardants on side-wall quenching of partially premixed laminar flames}

\author[stfs]{Matthias Steinhausen$^{1,*}$, Federica Ferraro$^{1}$, Max Schneider$^{1}$, Florian Zentgraf$^{2}$, Max Greifenstein$^{2}$, Andreas Dreizler$^{2}$, Christian Hasse$^{1}$, Arne Scholtissek$^{1}$}
\email{steinhausen@stfs.tu-darmstadt.de} 
\address[]{$^1$Technical University of Darmstadt, Department of Mechanical Engineering, Simulation of reactive Thermo-Fluid Systems, Otto-Berndt-Stra{\ss}e 2, 64287 Darmstadt, Germany\\
$^2$Technical University of Darmstadt, Department of Mechanical Engineering, Reactive Flows and Diagnostics, Otto-Berndt-Stra{\ss}e 3, 64287 Darmstadt, Germany
}

\begin{document}
\pagestyle{plain}

\maketitle

\begin{abstract} 
    A combined experimental and numerical investigation of partially premixed laminar methane-air flames undergoing side-wall quenching (SWQ) is performed. A well-established SWQ burner is adapted to allow the seeding of the main flow with additional gaseous products issued from a (secondary) wall inlet close to the flame's quenching point. First, the characteristics of the partially premixed flame that quenches at the wall are assessed using planar laser-induced fluorescence measurements of the \ce{OH} radical, and a corresponding numerical simulation with fully-resolved transport and chemistry is conducted. A boundary layer of enriched mixture is formed at the wall, leading to a reaction zone parallel to the wall for high injection rates from the wall inlet. Subsequently, in a numerical study, the wall inflow is mixed with dimethylmethylphosphonat (DMMP), a phosphor-based flame retardant. The DMMP addition allows the assessment of the combined effects of heat loss and flame retardants on the flame structure during flame-wall interaction. With an increasing amount of DMMP in the injected mixture, the flame stabilizes further away from the wall and shows a decrease in the local heat-release rate. Thereby, the maximum wall heat flux is significantly reduced. That results in a lower thermal load on the quenching wall. The flame structure analysis shows an accumulation of the intermediate species \ce{HOPO} at the wall similar to the \ce{CO} accumulation during the quenching of premixed flames without flame retardant addition. The study shows how the structure of a partially premixed flame is influenced by a wall that releases either additional fuel or a mixture of fuel and flame retardant. The insights gained from the canonical configuration can lead to a better understanding of the combined effects of flame retardants and heat losses in near-wall flames.
\end{abstract}


\section{Introduction} \addvspace{10pt}
Chemical inhibition of combustion processes is of considerable scientific and practical relevance for the prevention and suppression of fire hazards~\cite{Masri1994}. Fire retardant compounds are commonly used as additives to polymers to enhance their resistance to ignition and to reduce the flame spread without significantly affecting their properties~\cite{Green1996}. They contribute to the gas phase decomposition products with flame extinguishing effects but may also promote the char formation on the polymer surface, which acts as a protective coating~\cite{Schmitt2007}. The flame retardant (FR) addition reduces the ignition probability of the material and slows the fire spread, providing more time to escape or for fire suppression activities. Due to their advantageous properties, polymer materials with FR are largely used in construction, furniture, technical equipment, and transportation.

Previous scientific investigations of FRs were mostly related to the search for suitable chemical substances that meet certain criteria, such as flame inhibition, non-hazardousness, and climate friendliness. For example, \ce{CF3Br}, also known as Halon 1301, one of the most used FR, has been phased out after the Montreal Protocol in 1987 for its ozone-depleting potential. Several agents have been investigated to identify suitable Halon substitutes with a similar fire suppression efficiency. phosphorous-containing compounds (PCCs) are one of the most prominent groups of FRs, especially due to their low toxicity and high efficiency~\cite{Wang2021c}. 
The inhibition mechanism and effectiveness of gaseous FRs are usually investigated in canonical flames. Spherically expanding flames~\cite{Sikes2019},  co-flow diffusion flames \cite{Bouvet2016a,Bouvet2016b}, counterflow flames~\cite{MacDonald1999,MacDonald2001} and premixed flames~\cite{Korobeinichev2002,Korobeinichev2005,Rybitskaya2008} doped with PCCs have been experimentally investigated to analyze the combustion suppression capability of different PCCs. The ignition delay time has been also investigated in a heated shock tube by Mathieu et al.~\cite{Mathieu2018}. Numerical simulations of one-dimensional laminar flames have often been reported in these works to analyze the effects of FR on the flame structure and to validate kinetic mechanisms~\cite{Takahashi2019, Jayaweera2005a, Korobeinichev2005, Babushok2016} for phosphorous-containing mixtures. It has been observed that the effectiveness of PCCs as gas-phase combustion inhibitors varies widely with the flame type~\cite{Bouvet2016b}. 

In a fire scenario, the onset of the fire is usually at the surface of a burnable compound and, hence, in proximity to cold walls. In these configurations, the flames are additionally influenced by heat losses to the walls that affect the combustion chemistry. In recent years, detailed experimental and numerical investigations of the combustion chemistry during flame-wall interactions (FWIs) have been carried out in generic configurations. Experimental studies of head-on quenching and side-wall quenching (SWQ) flames were recently reviewed in~\cite{Dreizler2015}. Jainski et al.~\cite{Jainski2017} introduced an atmospheric SWQ burner that was extended and used in multiple experimental investigations~\cite{Kosaka2018, Kosaka2019, Zentgraf2021, Zentgraf2021a} of the thermochemical state and local heat-release rate (HRR) in laminar and turbulent flames. In the most recent studies, Zentgraf et al. measured the thermochemical state, represented by the temperature and the mole fractions of \ce{CO} and \ce{CO2} (measured simultaneously), in laminar~\cite{Zentgraf2021a} and turbulent~\cite{Zentgraf2021} flames. 
In addition to the experiments, corresponding numerical investigations with fully-resolved chemistry and transport were conducted. Ganter et al.~\cite{Ganter2017} investigated the \ce{CO} production mechanism close to the wall in a two-dimensional subdomain of the SWQ burner, and the authors were able to achieve very good agreement with the measurement results~\cite{Jainski2017}. In a similar numerical configuration, Steinhausen et al.~\cite{Steinhausen2021} assessed the local HRR in a dimethyl ether-air flame using the measurements performed in~\cite{Kosaka2019} as validation data. 
Palulli et al.~\cite{Palulli2019} assessed the \ce{CO} formation in laminar SWQ flames prone to velocity fluctuations using a similar setup. Further, the thermochemical state during turbulent FWI has been investigated in multiple direct numerical simulations in a fully developed channel flow~\cite{Gruber2010, Ahmed2021, Steinhausen2022}. 

Even though FWI and FR have been independently investigated in generic configurations in detailed numerical and experimental studies, only a few combined investigations of the effects of FR and heat losses on the near-wall flame structure are reported in the literature.
This study aims to narrow this gap by investigating a partially premixed methane-air flame mixed with dimethylmethylphosphonat (DMMP) undergoing SWQ in a combined numerical and experimental study.
First, the SWQ configuration investigated in~\cite{Zentgraf2021a} is extended by a secondary inlet at the wall that allows seeding of the main flow with gaseous products close to the flame's quenching point. 
This adaption of the burner results in a partially-premixed flame structure that is significantly different from the premixed flames investigated previously~\cite{Jainski2017, Kosaka2018, Kosaka2019, Zentgraf2021, Zentgraf2021a} and enables the assessment of the effect of an active wall releasing fuel or flame retardants towards the flame.
With a pure methane inflow at the wall, the behavior of the partially premixed flame during SWQ is assessed with planar laser-induced fluorescence (PLIF) measurements of the \ce{OH} radical and a corresponding numerical simulation. In a numerical study, the methane inflow from the wall is mixed with DMMP, a PCC, and the combined effects of FR and heat losses on the combustion chemistry are investigated.

\section{Experimental and numerical setup} \addvspace{10pt}

\begin{figure}[!htb]
    \centering
    \includegraphics[width=192pt]{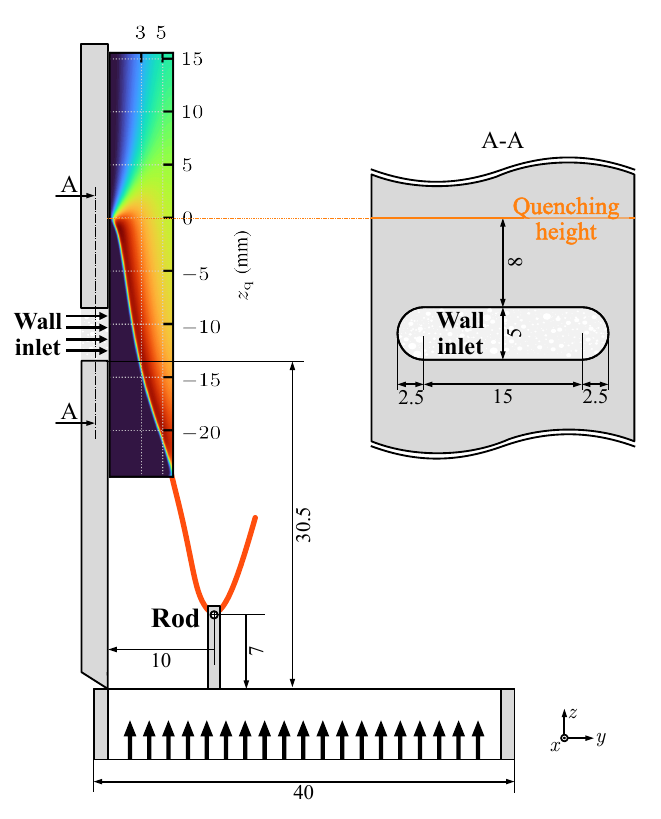}
    \caption{Schematic view of the SWQ burner and the numerical subdomain used for simulation. In section A-A (right), a slice through the wall is shown. The wall inlet is a porous media through which the methane(-DMMP) mixture is seeded to the main flow. The simulation and measurements are performed in a 2D subdomain that is located in the middle of the wall, i.e., in the middle of the porous media. All scales are given in mm.}
    \label{fig:setup}
\end{figure}

In the following, first, the experimental SWQ burner and the measurement setup are outlined. Secondly, the numerical setup is described. \figurename~\ref{fig:setup} shows a schematic view of the SWQ burner and the numerical subdomain investigated in this work. 
The burner configuration and measurement setup are similar to the one investigated in~\cite{Zentgraf2021a} with a modification of the quenching wall for additional fuel seeding.
The burner is operated at ambient conditions (around 293.15~K, 1~atm) and with laminar inflow. At the nozzle exit a fully premixed stoichiometric methane-air flow enters the domain at a Reynolds-number of $\text{Re} = u_\text{bulk}d_\text{h} / \nu \approx 5900$, with $u_\text{bulk}$ being the bulk velocity, $d_\text{h}$ the hydraulic diameter of the nozzle and $\nu$ the kinematic viscosity of the mixture. A V-flame is stabilized at a ceramic rod with a diameter of 1~mm. One flame branch is then quenched at the wall with a temperature stabilized at around 333~K. A porous media (sintered structure inlay) is placed inside the wall to allow a secondary inflow close to the flame's quenching point. The wall inlet is located 30.5~mm downstream of the nozzle exit and has a height of 5~mm. The detailed geometry is shown in the sectional drawing A-A in \figurename~\ref{fig:setup}. Through the porous media, methane at around 293.15~K is seeded to the premixed main flow, resulting in a partially premixed methane-air flame in the vicinity of the wall. The volume flow of methane at the wall is controlled by a mass flow controller. In the experimental campaign, the flame is visualized using PLIF measurements of the \ce{OH} radical, shown in \figurename~\ref{fig:validation}. 
For the PLIF measurements, the Q$_1$(9)+Q$_2$(8) line pair of OH was excited by pulsed UV-radiation around 284~nm using a frequency-doubled dye laser (pumped by a Nd:YAG at 532~nm, 10~Hz). The UV beam was formed to a light sheet with approximately 220~\textmu m thickness. The emission was collected by a UV lens (Sodern, Cerco 2073, 100~mm) equipped with a bandpass filter, and detected by an intensified relay optic (LaVision GmbH) coupled to a CCD camera (LaVision GmbH, Imager E-Lite 1.4M). For more details on the burner geometry and the measurement setup, the reader is referred to the partially premixed investigations performed in~\cite{Zentgraf2022}.

Following previous numerical investigations of the SWQ burner~\cite{Ganter2017, Steinhausen2021}, the numerical setup consists of a two-dimensional subdomain of the quenching flame branch.
The computational mesh is a rectilinear grid with a uniform grid size of 50~\textmu m. The computational domain has a length of 6~mm in the wall-normal direction and a height of 40~mm in the wall-parallel direction as shown in \figurename~\ref{fig:setup}. 
Similar to~\cite{Ganter2017, Steinhausen2021}, the inlet is modeled using a generic parabolic velocity profile. Hot exhaust gases are injected in a 0.5~mm wide section to stabilize the flame. In this area, the inflow velocity is increased by a factor of 2.244 to compensate for the density difference of the fresh and burnt gases. The top and right outlets are modeled using zero gradient boundary conditions for enthalpy, species, and velocity. For the pressure, a Dirichlet boundary condition is employed. The wall is modeled using a constant temperature of 333~K, a no-slip boundary condition for the velocity and for the species, a zero gradient boundary condition is employed. At the wall inlet (WI), a zero gradient boundary condition for the pressure is used, while the inflow velocity is assumed to be uniform over the inlet and therefore given by

\begin{equation}
    u_\text{WI} = \frac{\dot{V}}{A_\text{WI}} \ ,    
\end{equation}

\noindent with $A_\text{WI}$ being the total area of the wall inlet and $\dot{V}$ the volume flow through the wall inlet. The species are modeled with a Robin boundary condition that accounts for both diffusive and convective fluxes at the wall

\begin{equation}
    \dot{m}^{''}_i = \dot{m}^{''}_\text{tot} Y_i - \rho D_i \frac{\partial Y_i}{\partial y} \ ,
\end{equation}

\noindent with $\dot{m}^{''}_\text{tot}$ being the total mass flux per area, $\rho$ the density of the injected gas and $\dot{m}^{''}_i$, $D_i$ and $Y_i$ being the mass flux per area, the diffusion coefficient and the mass fraction of the species $i$ in the injected gas, respectively. The simulations are performed using an in-house solver with fully-resolved transport and chemistry (FTC) based on OpenFOAM (v2006). The species diffusion coefficients are modeled using a mixture-averaged transport approach~\cite{Curtiss1949, Coffee1981}, and the reaction mechanism of Babushok et al.~\cite{Babushok2016} is employed. Due to the minor effect on the results, radiative heat transfer is neglected in the simulations shown in the main part of the paper. For the interested reader, the radiation effects are discussed in the appendix. The results are plotted in a coordinate system relative to the quenching height. Two definitions of the quenching height are used: 

\begin{figure*}[!t]
    \centering
    \includegraphics[scale=1.0]{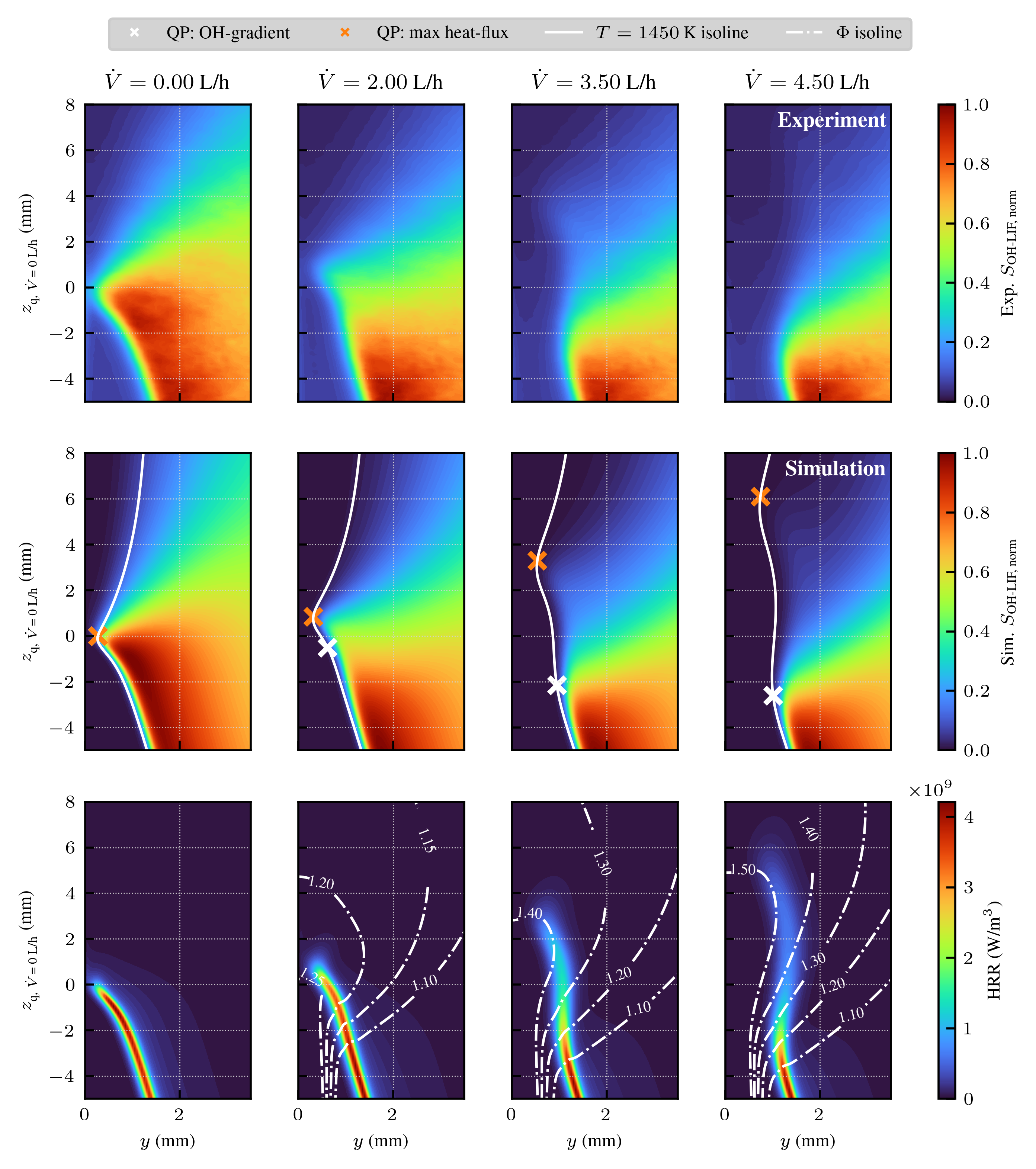}
    \caption{Normalized OH-PLIF measurements (top), computed OH-PLIF signals (middle) and local HRR (bottom) based on the numerical simulations. Different methane wall inflow rates (left to right) are shown, given in norm liter per hour. In the middle row, the 1450~K temperature isoline is shown together with two quenching point (QP) definitions, based on the OH-gradient and the maximum heat flux to the wall. In the bottom row, isolines of the local equivalence ratio are shown as dashed-dotted lines.}
    \label{fig:validation}
\end{figure*}

\begin{itemize}
    \item \textbf{OH-gradient:} The quenching height is based on the OH-gradient as used in~\cite{Zentgraf2021a, Ganter2017, Steinhausen2021}. This definition has been mainly used in experimental investigations, where the flame position was determined by qualitative OH-PLIF measurements.
    \item \textbf{Wall heat flux:} The quenching height is located at the maximum heat flux at the wall
    \begin{equation}
        \dot{q}_\text{wall} = - \lambda \left| \frac{\partial T}{\partial y} \right|_\text{wall} \ ,
    \end{equation}
    with $\lambda$ being the thermal conductivity and $y$ the wall normal direction. This has also been used in~\cite{Kosaka2018}. 
\end{itemize}

\noindent For both quenching height definitions, the quenching distance $y_\text{q}$ is then defined by the wall distance of the 1450~K temperature isoline at the respective quenching height~\cite{Kosaka2018}.

\section{Laminar side-wall quenching with methane addition} \addvspace{10pt}
\label{sec:validation}
Before the effect of flame retardants is discussed, the numerical setup for laminar SWQ with an additional \ce{CH4} wall inflow is validated utilizing OH-PLIF measurements. Furthermore, the characteristics of the partially premixed flame are discussed. \figurename~\ref{fig:validation} shows the measured normalized OH-PLIF signal (top, ensemble-average based on 400 samples) and its numerical counterpart (middle) that is calculated from the thermochemical state of the simulation, following~\cite{Steinhausen2021}. At the bottom of the plot, the local HRR of the flame is shown together with isolines of the local equivalence ratio (dashed-dotted lines) taken from the numerical simulations. From left to right different injection rates of \ce{CH4} are shown. Note that the wall inflow is very small in comparison to the inflow at the nozzle. For the maximum volume flow shown in \figurename~\ref{fig:validation} the ratio of the inlet velocities is  
\begin{equation}
    \frac{u_\text{WI}}{u_\text{bulk}} = \frac{0.014~\mathrm{ms^{-1}}}{2.12~\mathrm{ms^{-1}}} \approx 0.6~\% \ .
\end{equation}  \vspace{5pt}

The data are plotted relative to the quenching height based on the OH-gradient of the case without \ce{CH4} addition at the wall ($\dot{V}=0~\text{L/h}$). The experimental (top) and numerical (middle) data are in excellent agreement for all investigated inflow rates. 
The two quenching point definitions introduced above are shown as white (OH-gradient) and orange (maximum heat flux) markers in the middle row of \figurename~\ref{fig:validation}. While for the flame without an additional wall inflow, the quenching point definitions lead to similar quenching heights, they deviate with increasing methane inflow at the wall (left to right). That is a result of the enrichment of the flame close to the wall. The richer flames produce less \ce{OH} compared to the stoichiometric flame, resulting in a lowered OH-PLIF signal in the reaction zone of the flame. For these partially premixed flames, the quenching point definition based on the OH-gradient fails to predict the correct quenching location, since the flame is not only affected by heat losses to the wall, but also by the change in the local equivalence ratio. For low inflow rates ($\dot{V}=2.0~\text{L/h}$), the flame is only slightly affected, showing a slight increase in quenching height and distance. At higher inflow rates ($\dot{V}\geq3.5~\text{L/h}$), however, a concentration boundary layer of enriched mixture is formed, leading to a reaction front parallel to the wall at increased quenching height and distance.


\section{Effect of flame retardants on side-wall quenching flames} 
\begin{figure*}[!htb]
    \centering
    \includegraphics[scale=1.0]{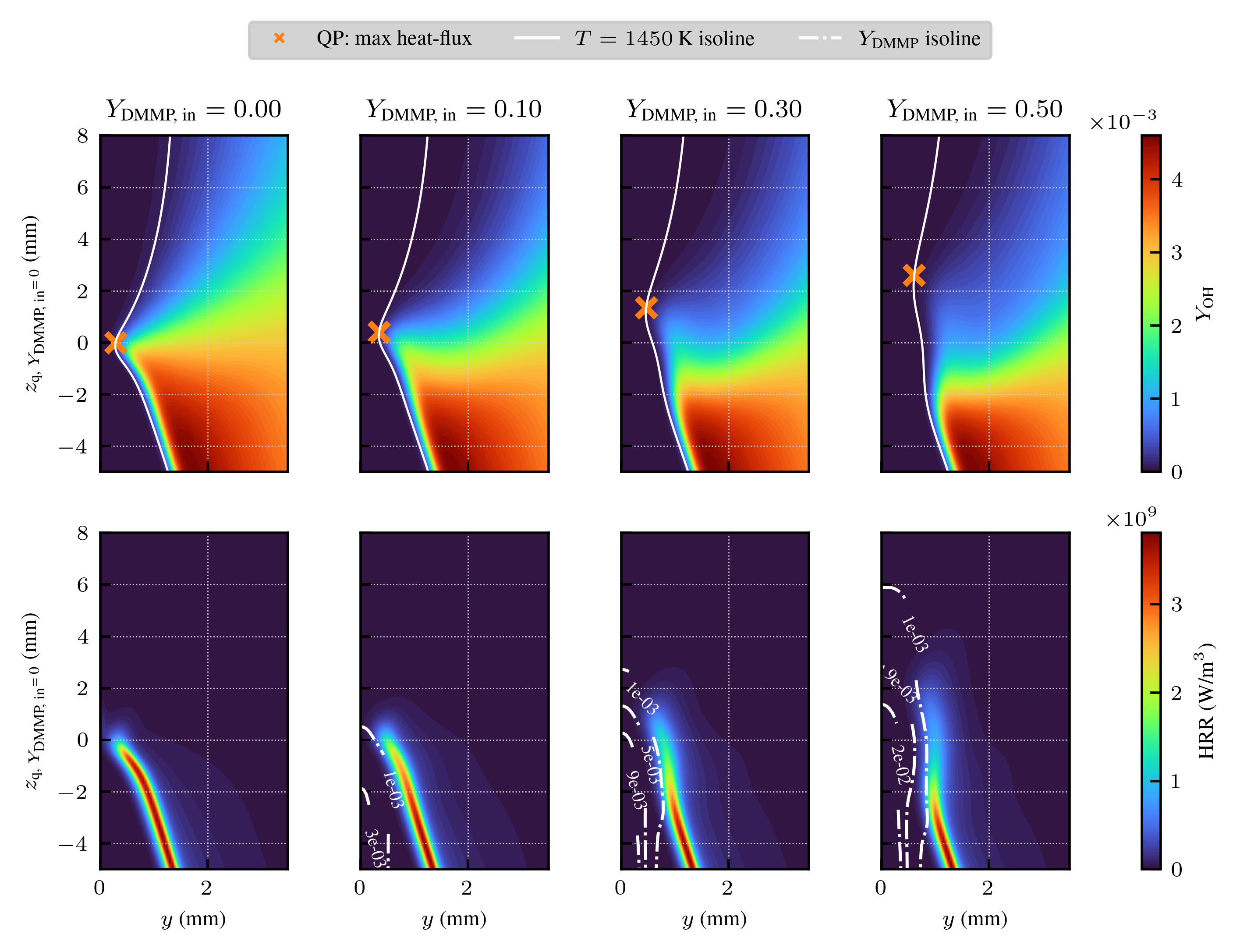}
    \caption{Mass fraction of \ce{OH} (top) and local heat-relase rate (bottom). From left to right the amount of DMMP in the injected mixture was increased. In the top row, the 1450~K temperature isoline is shown together with the quenching point based on the maximum heat flux to the wall. In the bottom row, isolines of the DMMP mass fraction are shown as dashed-dotted lines.}
    \label{fig:DMMP-variation}
\end{figure*}

\addvspace{10pt}
In the following, the effect of FR on the partially premixed methane-air flame is investigated.
The case with a wall inflow of 2.0~L/h is chosen as the base case for the analysis since it already shows an influence of the concentration boundary layer on the flame. 
However, compared to higher inflow rates, the flame shape and quenching distance of the flame are only slightly affected (see \figurename~\ref{fig:validation}), allowing the investigation of the combined effect of heat loss and FR at the wall. 
In this study, the methane inflow is gradually mixed with the flame retardant DMMP ($Y_\text{\ce{CH4}, in}=1-Y_\text{DMMP, in}$) while keeping the volume flow through the wall inlet constant. The inflow temperature is increased to 373.15~K to prevent condensation of DMMP. \figurename~\ref{fig:DMMP-variation} shows the simulation results for different compositions of the gas mixture injected at the wall inlet with an increasing amount of DMMP from left to right. Note that the amount of DMMP in the inlet mixture at the wall ($Y_\text{DMMP, in}$) is much larger than the amount of DMMP in the concentration boundary layer due to the low inflow velocity at the wall inlet ($u_\text{WI}/ u_\text{bulk}<0.3~\%$). For $Y_\text{DMMP, in}=0.5$ the maximum mass fraction of DMMP in the simulation domain is smaller than 0.1, which is further reduced when approaching the reaction zone of the flame (see the isolines in Fig.~\ref{fig:DMMP-variation}). The data are plotted relative to the quenching height based on the maximum heat flux of the case without DMMP addition. 

At the top of the figure, the mass fraction of \ce{OH} is shown together with the 1450~K temperature isoline and the quenching point based on the maximum wall heat flux for each case. At the bottom, the local HRR of the flame is shown together with isocontours of the DMMP mass fraction. Similar to the pure methane case, a boundary layer of DMMP develops at the wall, which changes the chemical composition of the burning mixture near the wall. This variation in mixture composition is decreasing the local HRR at the flame tip, which also affects the quenching at the wall. 

\begin{figure}[!htb]
    \centering
    \includegraphics[width=192pt]{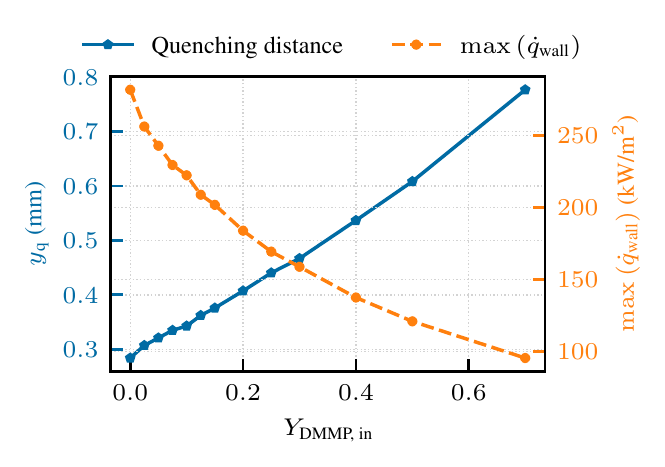}
    \caption{Quenching distance to the wall based on the maximum wall heat flux and the 1450~K temperature isoline. Additionally, the peak wall heat flux is shown.}
    \label{fig:xq-flux-DMMP}
\end{figure}

\figurename~\ref{fig:xq-flux-DMMP} shows the quenching distance and the maximum wall heat flux as a function of the mass fraction of DMMP in the gas mixture injected at the wall inlet. With an increasing amount of DMMP the quenching point moves downstream (see \figurename~\ref{fig:DMMP-variation}) and further away from the wall. At the same time, the maximum wall heat flux is decreased. This decrease can be explained by two phenomena: (i) close to the wall a non-flammable mixture is present due to the high amount of FR in the mixture; (ii) the enrichment of FR leads to a lower HRR and subsequently lower temperatures of the mixture with FR. 

\begin{figure}[!htb]
    \centering
    \includegraphics[width=192pt]{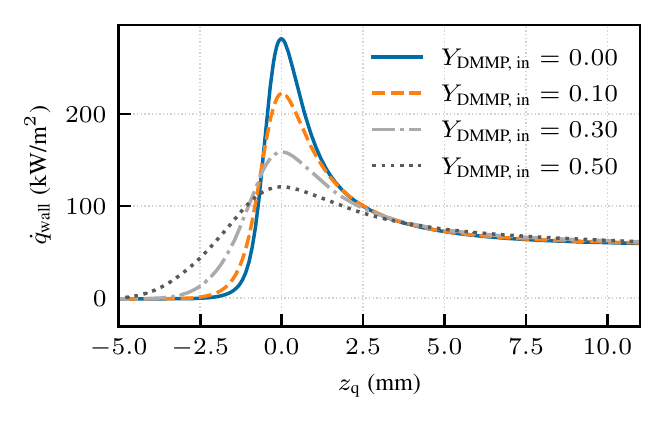}
    \caption{Wall heat flux over the wall-parallel coordinate for different inflow mixtures injected at the wall inlet.}
    \label{fig:heat-flux-DMMP}
\end{figure}

In addition to the maximum wall heat flux, \figurename~\ref{fig:heat-flux-DMMP} shows the heat flux profile over the wall-parallel coordinate relative to the quenching height. The addition of FR to the flame results in a significantly lowered peak heat flux at the wall, while the integral heat flux in the quenching region is approximately constant. The lowered peak heat flux to the wall reduces the thermal load on the wall and thereby the risk of fire outbreaks.

\begin{figure*}[!htb]
    \centering
    \includegraphics[scale=1.0]{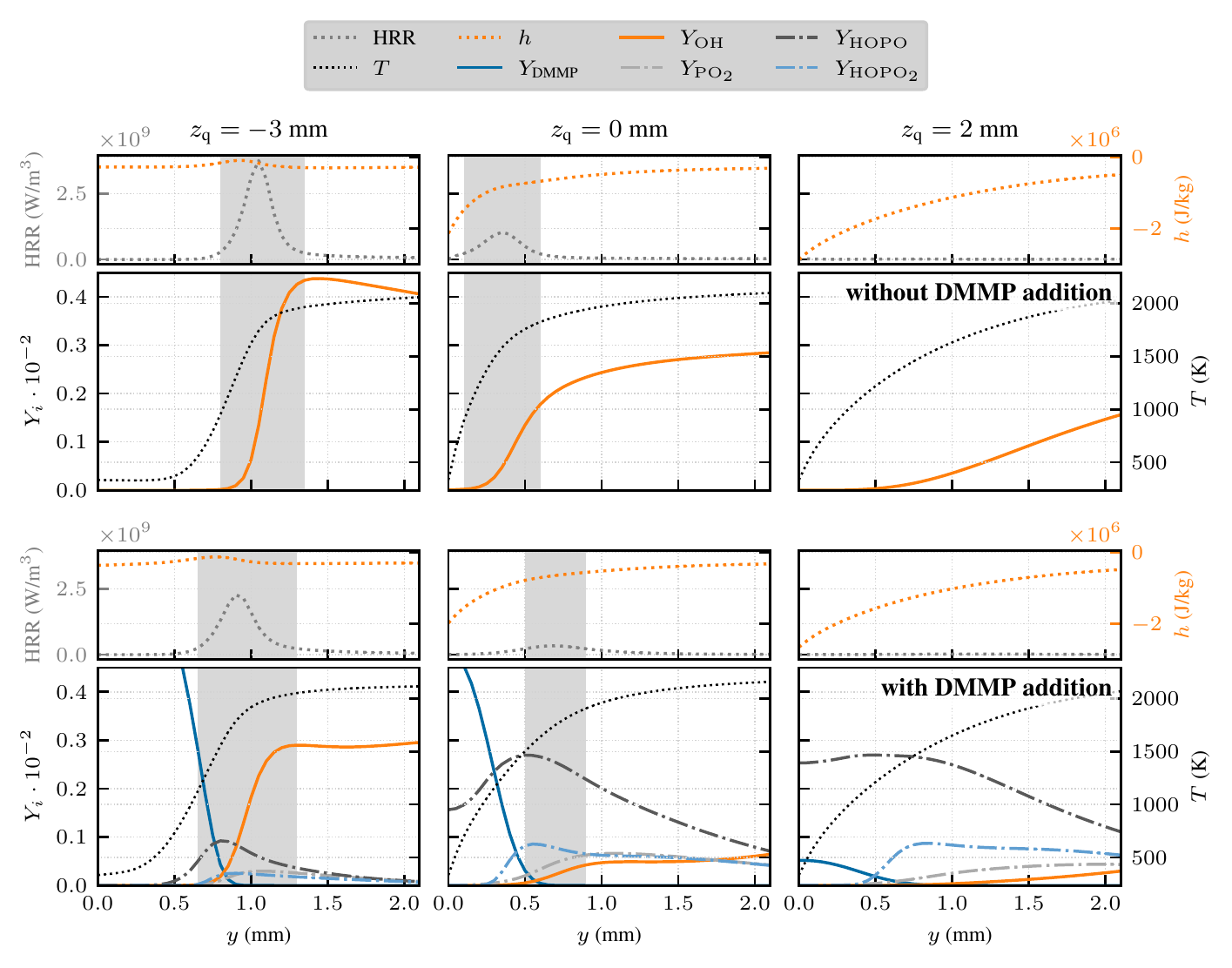}
    \caption{Flame structure at different stream-wise positions relative to the quenching point based on the maximum heat flux to the wall (orange marker in \figurename~\ref{fig:DMMP-variation}). The top panel corresponds to the case without DMMP addition at the inlet, while the lower panel corresponds to $Y_\text{DMMP, in}=0.3$. In the top subplots, the local HRR and the enthalpy are shown, while in the bottom the mass fractions of \ce{OH} and the most important phosphorous species are depicted. The grey box in the plots depicts the reaction zone of the flame ($\text{HRR}>2\cdot10^8~\text{W/m}^3$).}
    \label{fig:flame-structure}
\end{figure*}

Figure~\ref{fig:flame-structure} shows the flame structure at different stream-wise locations for a reference case without the addition of FR (top) and with DMMP addition (bottom), where a mixture of methane and DMMP is injected at the wall inlet. In the individual subplots, the local HRR and total enthalpy $h$ are shown in the top subplot, while at the bottom, the mass fractions of \ce{OH}, DMMP, \ce{PO2}, \ce{HOPO} and \ce{HOPO2} are depicted. 
Note that the \ce{H} radical shows a similar trend to the \ce{OH} radical and is therefore not shown for clarity in the figures.
The species were chosen following Babushok et al.~\cite{Babushok2016}, who identified these species as those involved in the most relevant reaction pathways of methane-air freely propagating flames diluted with PCCs. In their work, they also provided an illustrative pathway analysis for the reaction pathway of the phosphorous species.
First, the local HRR and the enthalpy loss to the wall (top subplots) are discussed for both flames. Then, the flame structure with DMMP addition is assessed in more detail using the mass fraction of \ce{OH} and the temperature profile of the flame without DMMP addition as a reference. 

In the region upstream the quenching point ($z_\text{q}=-3~\text{mm}$) both flames are not yet affected by enthalpy losses to the wall, showing an approximately constant enthalpy profile in the wall-normal direction $y$. In the flame with DMMP addition, the HRR already shows a lowered peak due to the enrichment of FR in the reaction zone. 
Once entering the reaction zone, the DMMP is consumed, \ce{HOPO} is formed and further converted to \ce{HOPO2} and \ce{PO2}, resulting in a lowered mass fraction of \ce{OH} and thereby inhibition of the flame. 
Note, however, that the phosphorous products are only present inside the reaction zone and the burnt part of the flame at higher temperatures and not in the near-wall region. 
At the quenching point ($z_\text{q}=0~\text{mm}$) both flames are affected by enthalpy losses to the wall, resulting in a lowered enthalpy in the near-wall region and a lowered HRR in both flames. Again, the flame with DMMP addition shows a significantly reduced HRR and mass fraction of \ce{OH}. It burns further away from the wall, resulting in a decreased temperature gradient and thereby a reduced heat flux to the wall (see \figurename~\ref{fig:heat-flux-DMMP}). 
A high amount of the DMMP is already consumed by the flame. In the reaction zone, the amount of the phosphorous species is increased compared to the upstream location. Similar to the upstream direction, \ce{PO2} and \ce{HOPO2} are only present in the high-temperature part of the flame. In contrast to that the intermediate phosphorous species \ce{HOPO} is built up at the wall. Finally, downstream the quenching point ($z_\text{q}=2~\text{mm}$) both flames are fully quenched. In this region, DMMP is almost entirely consumed, while \ce{HOPO} is (still) accumulated at the wall. A similar accumulation can also be observed for CO during laminar SWQ of fully premixed flames~\cite{Ganter2017}.

\section{Conclusion} \addvspace{10pt}
In this study, the effect of a (cold) wall, which either releases additional fuel or a mixture of fuel and flame retardant, on a partially premixed flame is assessed in a canonical side-wall quenching configuration. 
First, the effect of pure fuel addition is considered. Therefore, an established experimental SWQ burner~\cite{Zentgraf2021a} has been adapted to allow seeding of the main flow with gaseous components at the wall near the quenching point. The OH-PLIF measurements are compared to corresponding numerical simulations, which show a very good agreement. The local enrichment at the wall leads to complex partially premixed flames. These flames show an increased quenching distance with increasing inflow rates from the wall inlet and build up a wide reaction zone parallel to the wall at high injection rates.
Subsequently, the combined effect of flame retardants and heat losses to the wall is assessed by adding the phosphor-based flame retardant DMMP to the mixture injected at the wall inlet. The addition of DMMP leads to lower local heat-release rates in the flame, which result in higher quenching distances and reduced maximum wall heat fluxes, i.e., a reduced thermal load. Further, the analysis of the structure of the flame with DMMP addition at the wall inlet shows an accumulation of intermediate phosphorous product species (\ce{HOPO}) in the near-wall region. A similar observation has been made for \ce{CO} in laminar flames undergoing SWQ without flame retardant addition. The insights gained from this canonical configuration lead to a better understanding of the effect of flame retardants on near-wall combustion, which is of relevance to fire safety and prevention. 
Future work should address the influence of flame retardants on turbulent boundary layer flames, which could be realized by an extension of the present active wall configuration. Together with suitable experiments, numerical simulations of such configurations could lead to important insights in the field of fire safety.


\section*{Appendix: Impact of radiative heat transfer} 
\begin{figure}[!htb]
    \centering
    \includegraphics[width=192pt]{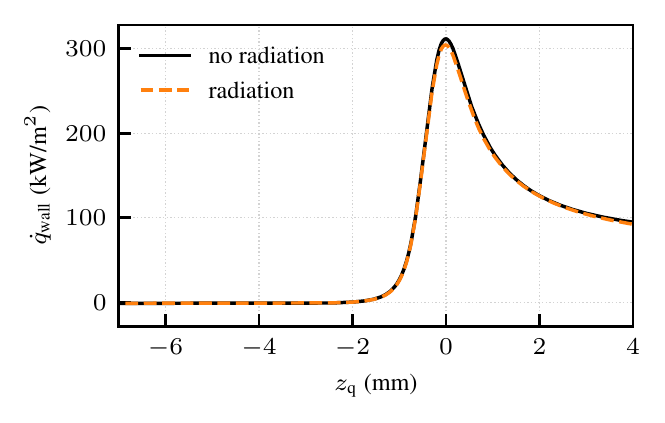}
    \caption{Wall heat flux over the wall-parallel coordinate with and without accounting for radiative heat transfer for the case without additional wall inlet ($\dot{V}=0.0$ L/h).}
    \label{fig:heat-flux-CH4-radiation}
\end{figure}

\addvspace{10pt}
To assess the role of radiation, the case without the additional wall inlet ($\dot{V}=0.0$~L/h) is simulated with and without modeling radiative heat transfer. The radiation model employed is described in~\cite{Barlow2001}. For the optically thin model (OTM), the radiative heat loss between a given fluid element in the flame and the cold surroundings is given by
\begin{equation}
    \dot{Q} = 4 \sigma \cdot \left( T^4 - T_b^4 \right)  \sum_i \left( p_i a_{p,i} \right) \ ,
\end{equation}
with $\sigma=5.669e^{-8}~\mathrm{W/m^2K^4}$, $T$ and $T_b$ being the local flame temperature and the background temperature, and $p_i$ and $a_{p,i}$ being the partial pressure and the Planck mean absorption coefficient of species $i$, respectively. The most important radiating species for hydrocarbon flames, i.e. H$_2$O, CO$_2$, CO and CH$_4$, are accounted for and their mean Planck absorption coefficients are calculated using RADCAL~\cite{Grosshandler1993}.
Figure~\ref{fig:heat-flux-CH4-radiation} shows the wall heat flux with and without radiative heat transfer. The inclusion of radiative heat transfer has only a minor effect on the results, which is also confirmed in the maximum heat flux, quenching distance, quenching height, and overall flame structure (not shown here). These findings are in accordance with the typical modeling assumption in flame-wall interactions that ”radiative fluxes are neglected [...] because they are small compared to the maximum heat flux obtained when the flame touches the wall”~\cite{Poinsot2005}.
In the flames diluted with DMMP, the amount of phosphorous species is very small, i.e. the elemental mass fraction is smaller than 1\% in the reaction zone of the flame. For this reason, the radiation of the phosphorous species is not expected to be of high relevance due to the low amount of phosphor in the flame. However, the information on the radiative properties of the phosphorous species is very sparse and their role in radiative heat transfer could be explored in future experimental studies.

\section*{Acknowledgments} \addvspace{10pt}
This work has been funded by the Deutsche Forschungsgemeinschaft (DFG, German Research Foundation) -- Project Number 237267381 -- TRR 150. Calculations were conducted on the Lichtenberg high-performance computer of the TU Darmstadt. We thank P. Johe from RSM for his support during the measurements.

\bibliography{publication.bib}
\bibliographystyle{unsrtnat_mod}

\end{document}